\documentclass{JINST}
\let\ifpdf\relax

\usepackage{graphicx}

\title{Measurement of Optical Attenuation in Acrylic Light Guides for a Dark Matter Detector}

\author {M. Bodmer$^a$, N. Phan$^a$, M. Gold$^a$, D. Loomba$^a$, J.A.J. Matthews$^a$ and K. Rielage$^{a,b}$\thanks{Corresponding
author.}\\
\llap{$^a$}University of New Mexico, Department of Physics and Astronomy,\\
  Albuquerque, NM 87131, USA\\
\llap{$^b$}Los Alamos National Laboratory,\\
  Los Alamos, NM 87545, USA\\
  E-mail: \email{rielagek@lanl.gov}}

\abstract{Acrylic is a common material used in dark matter and neutrino detectors for light guides, transparent vessels, and neutron shielding, creating an intermediate medium between the target volume and photodetectors.  Acrylic has low absorption within the visible spectrum and has a high capture cross section for neutrons.  The natural radioactivity in photodetectors is a major source of background neutrons for low background detectors making the use of acrylic attractive for shielding and background reduction.  To test the optical properties of acrylic we measured the transmittance and attenuation length of fourteen samples of acrylic from four different manufacturers.  Samples were evaluated at five different wavelengths between 375 nm and 632 nm.  We found that all samples had excellent transmittance at wavelengths greater than 550 nm.  Transmittance was found to decrease below 550 nm.  As expected, UV-absorbing samples showed a sharp decrease in transmittance below 425 nm compared to UV-transmitting samples.  We report attenuation lengths for the three shortest wavelengths for comparison and discuss how the acrylic was evaluated for use in the MiniCLEAN single-phase dark matter detector.}


\begin{document}

\section{Introduction}
\large\normalsize

\begin{figure}\centering
\includegraphics[width=.6\linewidth]{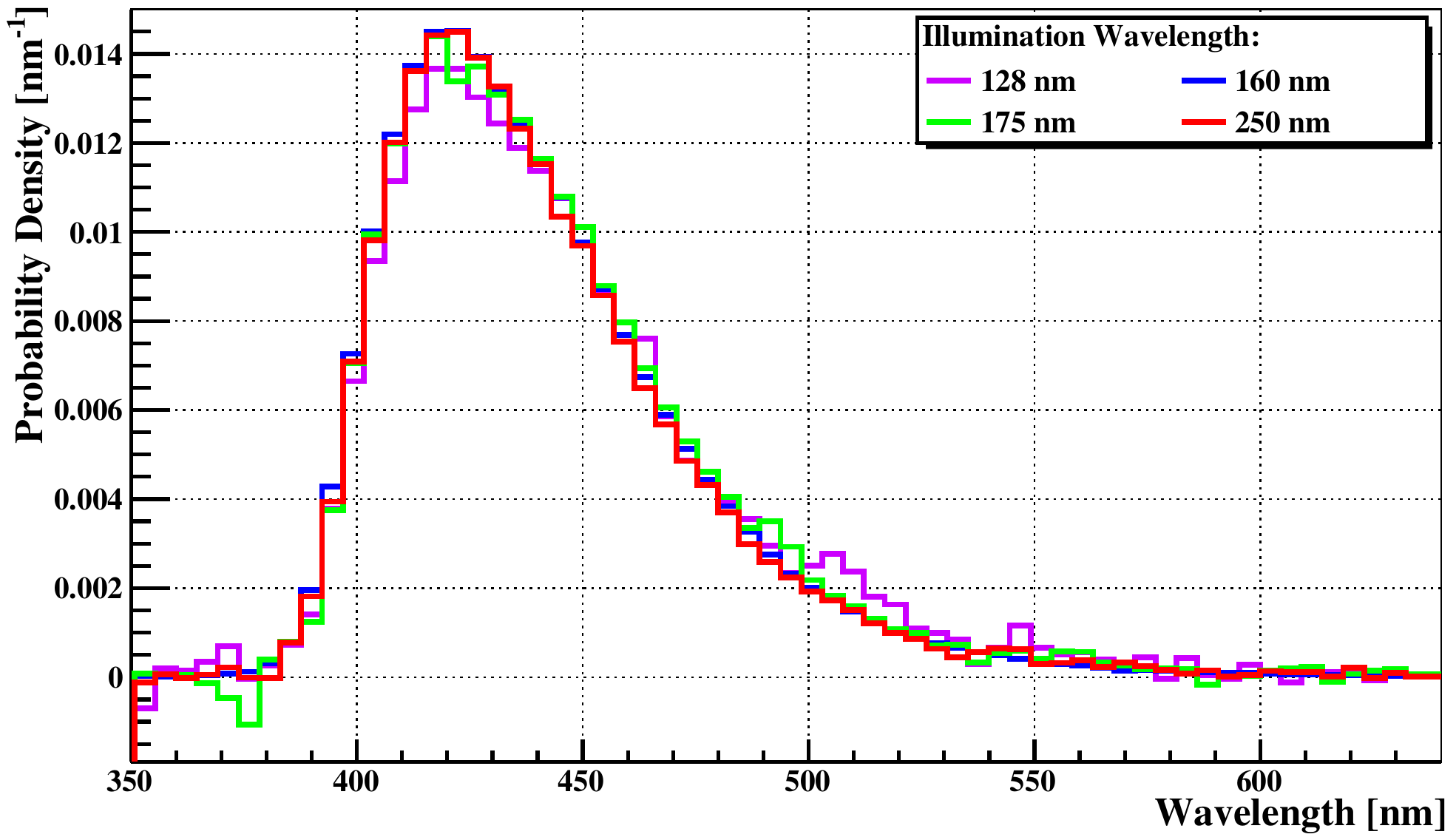}
	\label{emission}
	\caption{The re-emission spectrum of TPB at four different excitation wavelengths.  All spectra have a mean emission at approximately 440 nm with a maximum at 420 nm. From \cite{bibTPB}.}
\end{figure}

Acrylic, or polymethyl methacrylate (PMMA), is often used by neutrino and dark matter detectors as an inexpensive light guide (e.g., LBNE \cite{lbne}), transparent containment vessels (e.g., SNO \cite{sno} and Daya Bay \cite{daya}\cite{daya-acrylic}), and for neutron shielding (e.g., DEAP-3600 \cite{deap}).  Acrylic is often chosen because it is a strong material and can be manufactured with low bulk radioactivity \cite{LRT} and high optical transmission. The acrylic samples in this paper were tested for use in the MiniCLEAN experiment \cite{miniclean}: a single-phase noble liquid WIMP dark matter detector.  The purpose of acrylic in MiniCLEAN is threefold. First, acrylic acts as a neutron absorber, capturing neutrons from natural radioactivity in the detector components before they reach the central detection volume.  This is important for MiniCLEAN since the natural radioactivity of the borosilicate glass in the photomultiplier tubes (PMTs), used to detect the scintillation light signals, generates approximately 40,000 neutrons per year. Second, the scintillation light produced by interactions in the liquid argon (LAr) and liquid neon (LNe) target is strongly peaked at 128 nm and 80 nm, respectively.  Acrylic provides a surface that can be coated with a wavelength shifter to convert the VUV scintillation light into the visible spectrum.  This surface defines the central active volume of the target.  For MiniCLEAN, tetraphenyl butadiene (TPB) has been selected as the wavelength shifter since its re-emission spectrum peaks at about 420 nm (shown in Figure \ref{emission}) and is well matched to the response of the bialkali photocathode on the PMTs.  The third function of the acrylic is as a light guide, creating a medium for the visible light emitted from the TPB coated surface to propagate through while being directed into the PMTs.  

\begin{figure}[htbp]
\begin{center}
\includegraphics[width=0.9\linewidth]{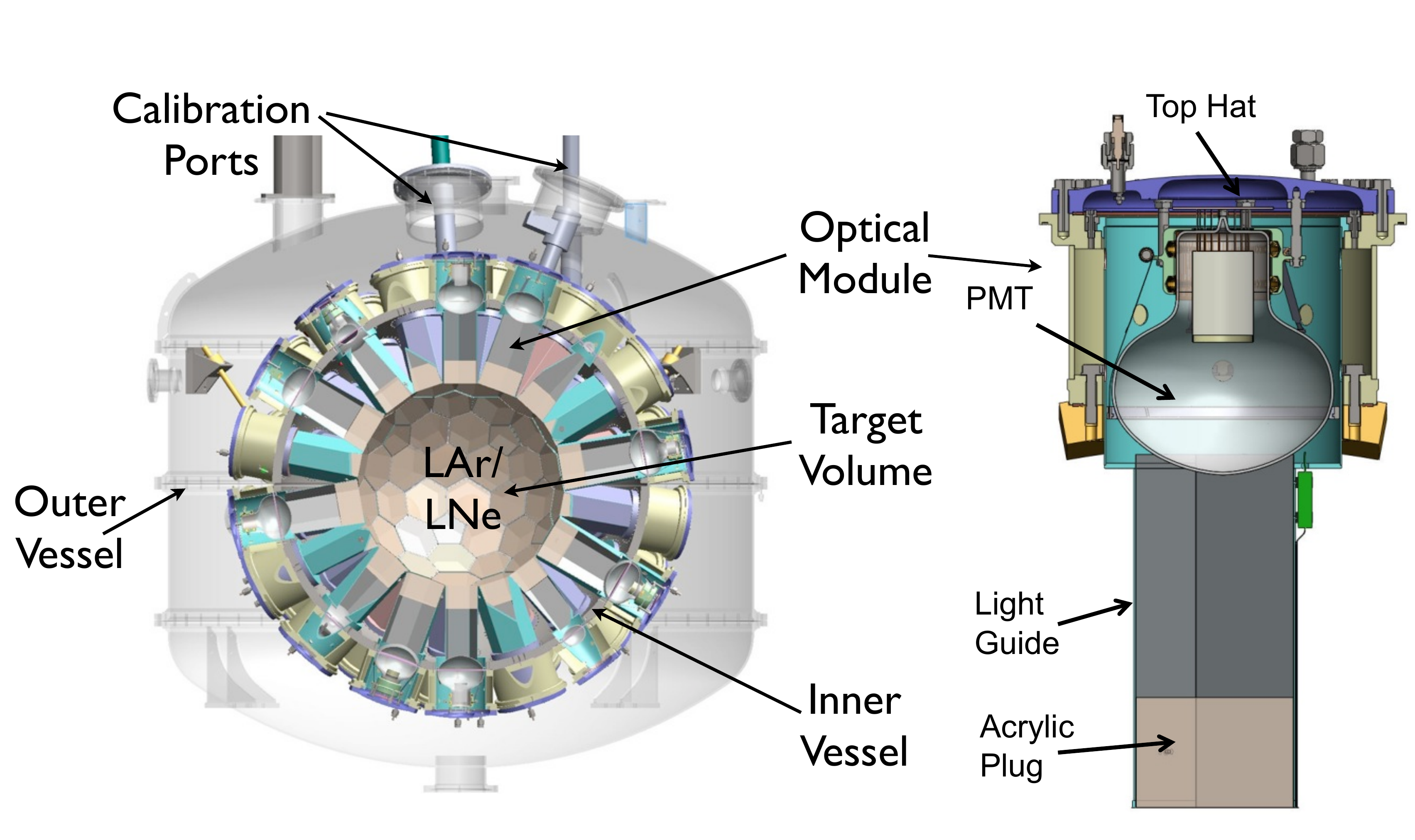}
\caption{A cutaway model of the MiniCLEAN detector and an optical cassette module.}
\label{MCcut}
\end{center}
\end{figure}

A model of the MiniCLEAN detector is shown in Figure \ref{MCcut}.  MiniCLEAN utilizes ninety-two optical cassette modules that each contain acrylic coated with TPB on the front surface and a Hamamatsu R5912-02MOD 8'' photomultiplier tube.  These optical cassette modules are inserted into a stainless steel inner vessel that contains the cryogenic target.  Monte Carlo simulations of the detector predict that approximately 6 photoelectrons will be detected per keV of electron equivalent energy.  The simulations show that in the 10 cm of acrylic approximately 6\% of the light will be lost due to attenuation and internal reflection losses.  The remaining light is lost from the conversion of the ultraviolet light to visible, the reflection losses along the light guide and on non-reflective parts of the detector, the photodetector efficiency and some scattering in the liquid argon.  Poor optical quality acrylic would reduce the overall light yield of the detector and affect the low energy threshold that could be obtained and dramatically reduce the physics capability of the detector.  In addition, a dark matter experiment is searching for a few signal events per year amid a large number of background events.  Thus, these experiments must require stringent limits on any materials that could produce background events that may be mistaken for signals.  Acrylic and a liquid argon buffer region greatly reduce the 40,000 neutrons produced in the photomultiplier array each year and when combined with data reduction cuts should result in 0.5 an event remaining in the data in the energy region of interest.  For acrylic near the active region of the detector in MiniCLEAN, this requirement places limits on alpha emitters from the radon chain that can be in the material.  To reduce this background material on the surface with the wavelength shifter is removed immediately prior to the deposition.  This removes radon and daughter nuclei that have diffused into the acrylic.  A bulk radioactivity of 1 mBq/kg in the acrylic will result in a background after data reduction cuts that is significantly less than 0.1 event per year in the energy region of interest.

The focus of this paper is to examine the optical response of acrylic as a light guide for wavelengths in the visible spectrum located around the emission peak of TPB at 420 nm.  We compare optical properties between samples and manufacturers to aid in decisions involving the selection of appropriate acrylic for future detectors and examine the specific factors that motivated the choice of acrylic manufacturer for the MiniCLEAN detector.  We are most concerned with the wavelengths between 380 nm and 500 nm where the transparency of acrylic differs greatly from 100\%.  Our measurement method has accounted for many systematic effects including positional dependence effects from non-uniform acrylic manufacturing, laser instability, photodetector variations, reflection variations on surfaces of the acrylic, background light, and sample to sample variation.

It is important to note that each manufacturer of acrylic utilizes different modifications and additives to adjust the properties of the resulting product.  A thorough discussion of the manufacture of acrylic can be found in \cite{handbook}.   There are three main methods for producing sheets, rods and shapes during the bulk polymerization process: cell casting, continuous casting and slush casting.  Cell casting is done by filling a cell between two pieces of tempered glass with the monomer-polymer mixture and sealing it with a gasket.  The material is then polymerized by placing the cell in water tanks, air ovens or autoclaves for a specified temperature and time.  A continuous process utilizes two stainless steel conveyor belts where the mixture is squeezed vertically between the two belts.  The top and bottom belt move at identical speeds and move the material through temperature zones of varying lengths and a cured sheet is produced at the end.  Slush casting is used to produce large or unusual shapes.  The polymer-catalyzed monomer mixture is poured into a mold and allowed to cure.  The mold can be stationary or rotary.  For this work, samples from cell casting and slush casting were tested.
 
One of the primary purposes of the additives in the monomer mixture is to tune the amount of ultraviolet light (below 380 nm) that is transmitted through the acrylic.  Material can be classified as ultra-violet absorbing (UVA), ultra-violet transmitting (UVT), or even super-ultra-violet transmitting (SUVT) depending on the mixture.  The samples in this study were limited to UVA and UVT but two samples that were a result of mixing an incomplete set of additives creating a UVA/UVT mix are included for the comparison.

\section{Experimental Set-up}
\begin{figure}[htbp]
\begin{center}
\includegraphics[width=0.7\linewidth]{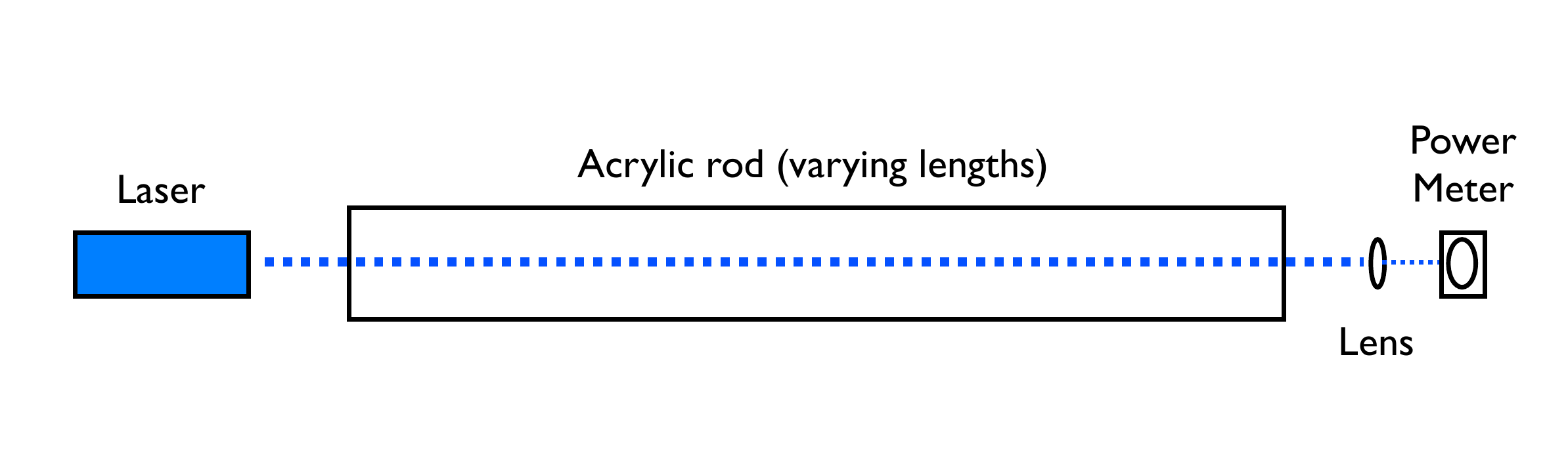}
\caption{Basic experimental setup where the light from a laser is measured with a power meter to study the light loss after traversing an acrylic sample.}
\label{setup}
\end{center}
\end{figure}

This section describes the experimental setup for measuring the optical attenuation of the acrylic samples.  The basic setup is shown in Figure \ref{setup}.  Lasers of differing wavelength are used as a stable monochromatic light source and a power meter is used to determine the amount of light lost as the beam traverses an acrylic sample.  This section provides details on the acrylic samples used and their preparation, the use of the power meter and lasers, and the data collected.  This setup has the disadvantage of only testing the optical quality of the acrylic at certain wavelengths but it improves the accuracy of the measurements by using long ($\approx$ meter) samples.

\subsection{Acrylic Samples}
Fourteen acrylic samples were purchased from four different sources:  Plastifab Tecacryl$^{\textregistered}$  \cite{Tecacryl}, Spartech\cite{spartech} (now PolyOne), McMaster-Carr \cite{carr} (McMaster), and Reynolds Polymer Technology R-Cast$^{\textregistered}$ \cite{rpt} (RPT).  Twelve of these samples had a cylindrical geometry while the remaining two were rectangular bars.  In addition samples are specified as UVA or UVT by the manufacturer.  For two of the Tecacryl samples an error in the manufacturing resulted in an incomplete mixture of the additives in the rods (labelled UVA/UVT).  Results from these samples are included to demonstrate the effect of such mixtures.  The samples from McMaster-Carr are part number 8581K86 (manufactured by Spartech) and 8528K47 (manufactured by US Cast Inc.). The sample descriptions are summarized in Table \ref{acrylic}. 

	\begin{table} 
	\begin{center}
	\begin{tabular}{|c|c|c|c|c|} \hline\hline
	{\em Manufacturer} & {\em Sample Number} & {\em Length (cm)} & {\em Diameter (cm)} & {\em Type} \\ \hline
	Tecacryl & \#1 & 94.11 & 20 & UVA/UVT \\
	Tecacryl & \#2 & 94.11 & 20 & UVA/UVT \\
	Tecacryl & \#3 & 94.11 & 20 & UVT \\
	Tecacryl & \#4 & 94.11 & 20 & UVT \\
	RPT & \#1 & 94.11 & 10.5 & UVT \\
	RPT & \#2 & 94.11 & 10.5 & UVT \\
	RPT & \#3 & 94.11 & 10.5 & UVT \\
	RPT & \#4 & 94.11 & 22.86 & UVA \\
	Spartech & \#1 & 94.11 & 20 & UVT \\
	Spartech & \#2 & 94.11 & 20 & UVT \\
	Spartech & \#3 & $\dagger$ & $\dagger$ & UVA \\
	Spartech & \#4 & 121.92 & $\ddagger$ & UVA \\
	McMaster (Spartech)  & \#1 & 15.24 & 20 & UVT \\
	McMaster (US Cast) & \#2 & 30.48 & 10.5 & UVT \\
	 \hline\hline
	\end{tabular}
	\end{center}
	\caption {Columns 1 \& 2 list the acrylic samples by their manufacturer and the sample number they received in the lab.  Columns 3 \& 4 list the dimensions of the acrylic assuming cylindrical symmetry.  All measurements were taken along the sample length.  The last column lists the type of acrylic as specified by the manufacturer.  $\dagger$ This sample did not have cylindrical geometry and was measured through multiple lengths.  Its dimensions are 30.48 cm $\times$ 30.48 cm $\times$ 10.16 cm.  It was measured through both 30.48 cm and 10.16 cm lengths. $\ddagger$ This sample did not have cylindrical geometry.  Its dimensions are 121.92 cm $\times$ 22.86 cm $\times$ 10.8 cm.  It was measured through its 121.92 cm length. }
	\label{acrylic}
	\end{table}

\subsection{Acrylic Rod Preparation}
Once the acrylic samples were received they were sent for machining at the University of New Mexico Physics and Astronomy Department shop.  Cylindrical pieces of acrylic were machined on a lathe using a series of rollers and steady rests to ensure proper alignment.  The desired face was then machined creating a uniform surface perpendicular to the axis of symmetry.  The sample was then rotated and the second face was machined in a similar fashion.  The machined surfaces were then wet sanded using 600 and 1200 grit sandpaper, finished using acrylic polish, and covered with masking tape for protection.  Rectangular samples were surfaced using a milling machine and an adjustable height die cart for stabilization.  After machining, the surfaces of the rectangular samples were polished and finished in the same manner as the cylindrical samples.

\subsection{Lasers and Power Meter}
Five lasers were utilized (listed in Table \ref{laser}) based on their output wavelength, to ensure that a broad selection of expected TPB emission wavelengths could be evaluated. The three lasers produced by Power Technology were solid state devices and were found to have very stable power outputs ($\approx 1\%$ fluctuation) and short warm up times. These lasers were allowed $\approx 30$ min of warm up time before any measurements were taken.  Neutral density filters were used to reduce the power output of these solid state lasers to a Class 1 level.  The remaining two lasers (543 nm and 632 nm) were Helium-Neon lasers and were found to have greater instability, $\approx 3\%$ for the 543 nm laser and $\approx 7\%$ for the 632 nm laser, and extremely long warm up times. These lasers required a warm up time of $\approx$ 3 hours.  The lasers were focused on a photo-detector (Newport Model 918-UV $\cite{newport}$) that was integrated with a power meter (Newport Power Meter, Model 1930c $\cite{newport}$).  The power meter was designed to measure the power of incoming light at a single wavelength which could be set to correspond with the emission wavelength of the mounted laser.

\begin{table} \centering
\begin{tabular}{|c|c|c|c|} \hline\hline
{\em Manufacturer} & {\em Model} & {\em Wavelength (nm)} & {\em Max Power Output (mW)} \\ \hline
Power Technology & PPMT 16 & 375 & 16 \\
Power Technology & PPMT 50 & 405 & 50 \\
Power Technology & LDCU 12/6415 & 440 & 3 \\
Research Electro-Optics Inc. & 30972 & 543 & 2 \\
Spectra-Physics Inc. & 155 & 632 & 0.95  \\
\hline \hline
\end{tabular}
\caption{List of the five lasers used and their specifications.  The first three lasers were solid state while the remaining two were helium-neon.}
\label{laser}
\end{table} 

\subsubsection{Positional Dependence of the Power Meter}
\begin{figure}\centering
\includegraphics[width=.49\linewidth]{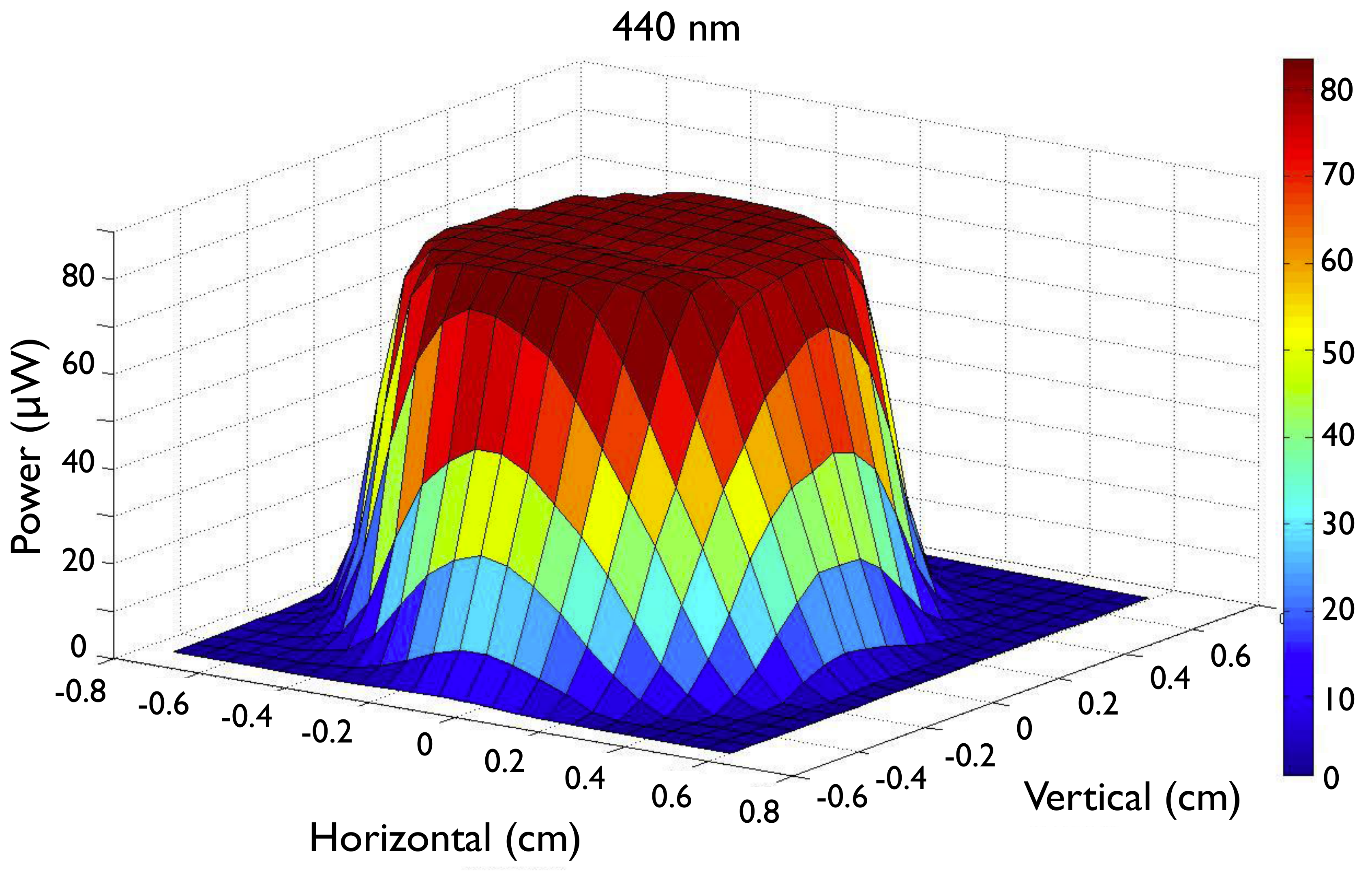}
\includegraphics[width=.49\linewidth]{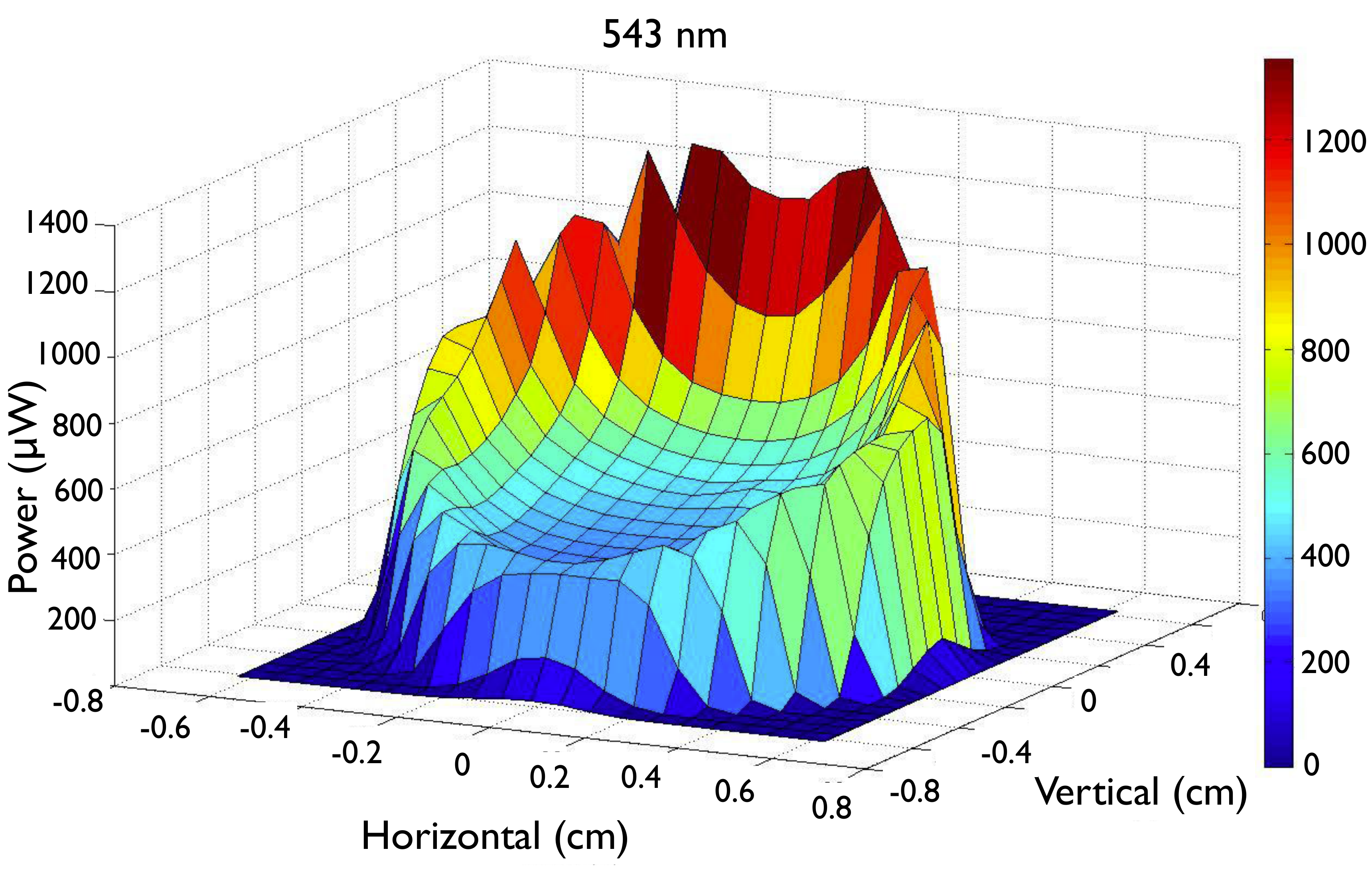}
\includegraphics[width=.49\linewidth]{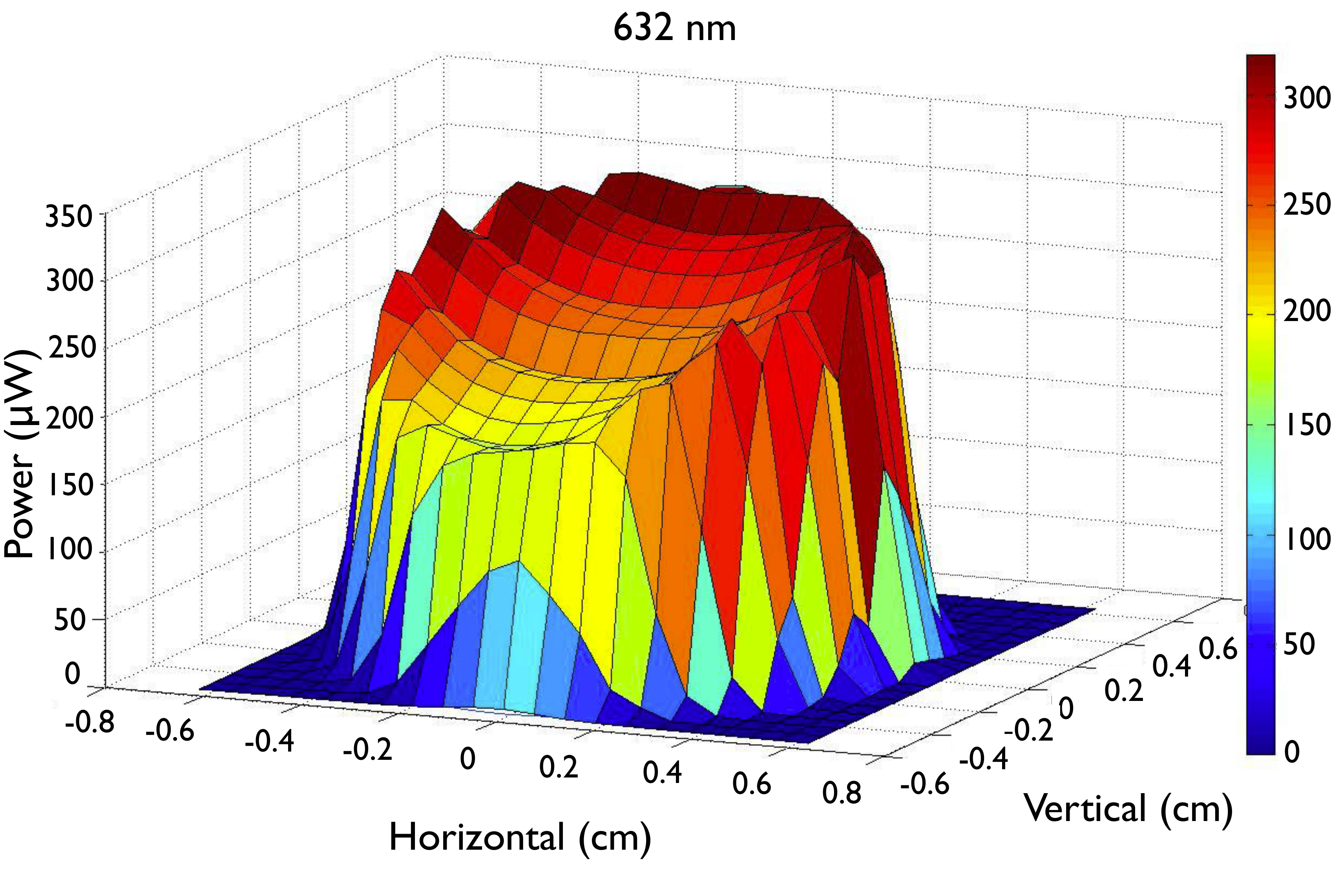}
\caption{Detector profile for 440 nm (upper left), 543 nm (upper right), and 632 nm (bottom) lasers.  The origin lies at the center of the detector meaning that points further away on the vertical axis represent the top of the detector. Measurements using the 405 and 375 nm lasers had similar profiles to the 440 nm laser shown above.}
\label{profile}
\end{figure}

\begin{figure}\centering
\includegraphics[width=.48\linewidth]{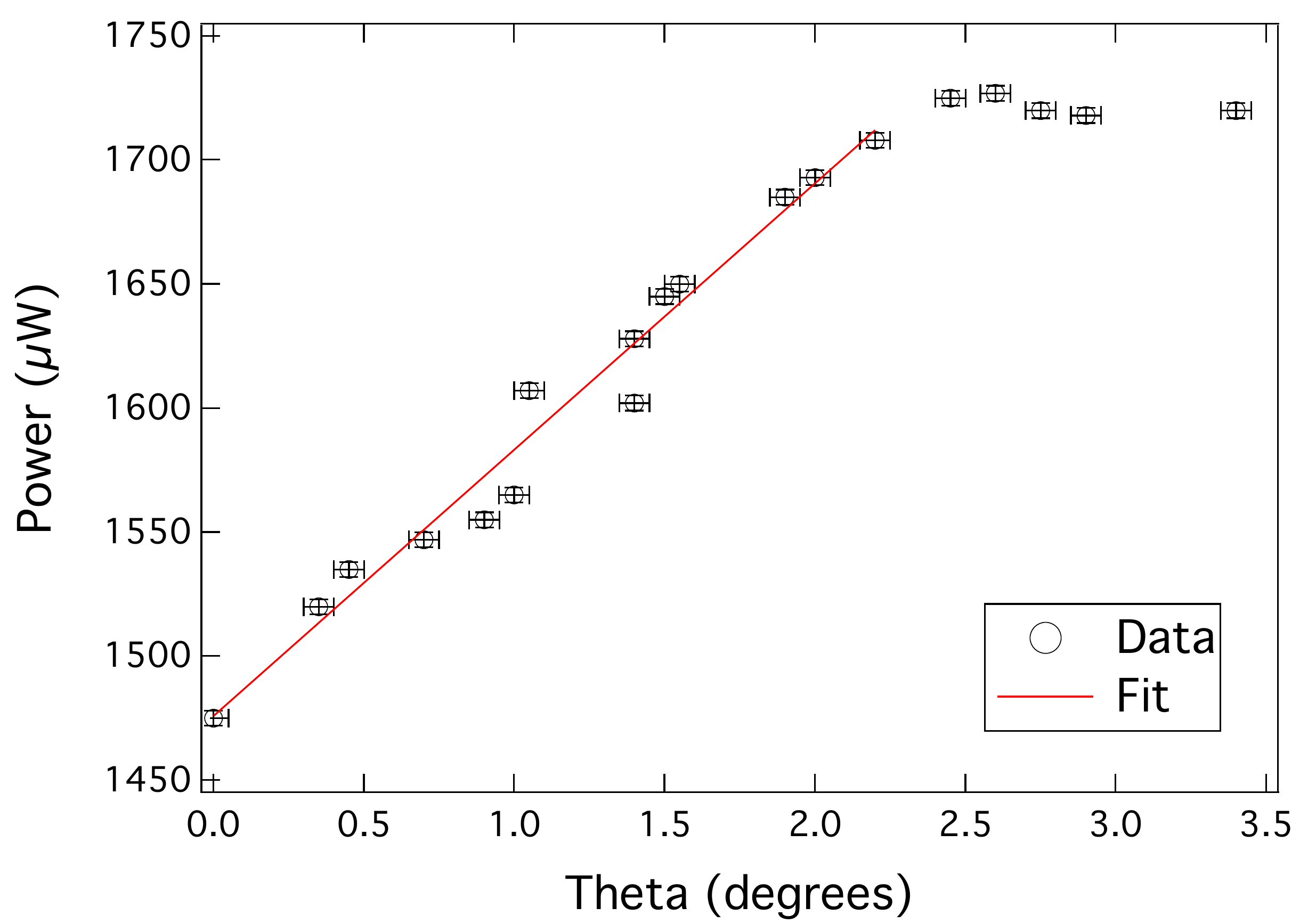}
\includegraphics[width=.46\linewidth]{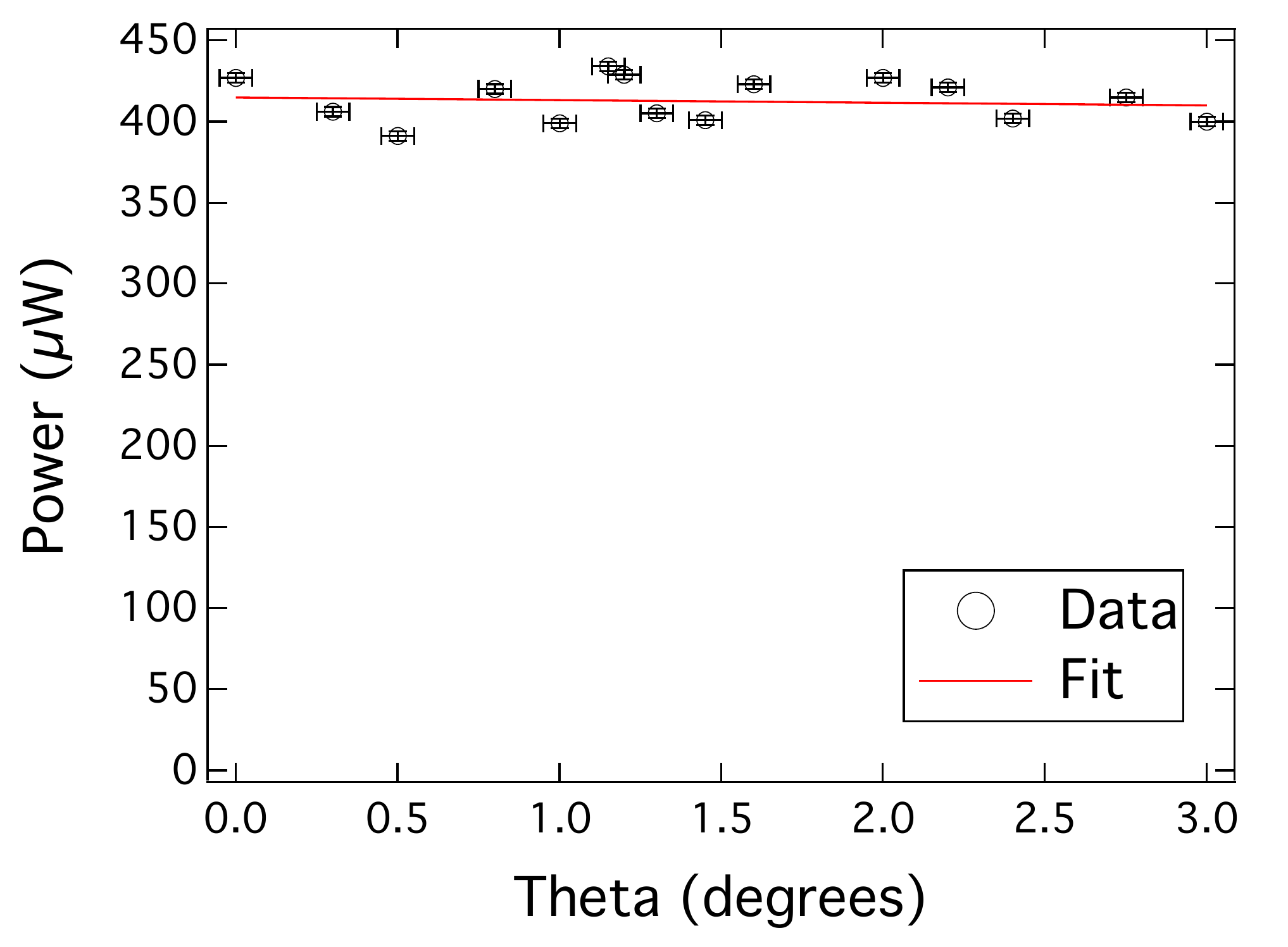}
\caption{Left: Measured power versus angle of light incident on the top edge of the detector.  Power readings rise as the angle of incidence is increased and plateau at angles greater than $\approx$ 2.25$^{\circ}$ .  Right: Measured power versus angle of light incident on the center of the detector.  Power readings remain relatively consistent as the angle of incidence is varied.  Both plots were made using the 543 nm laser.}
\label{angle}
\end{figure}

During initial testing of the setup a systematic error was observed when measuring the power of the 543 and 632 nm lasers.  Occasionally the recorded power of the laser passing through the acrylic sample would be larger than the power measured of the laser through air.  These results directly contradict the idea that the acrylic attenuates the beam suggesting that there was a problem with the detector.  By aiming the 543 nm laser at different sections of the photodetector a large variance was observed between measurements taken at the center and those at the edges. This effect was duplicated with the 632 nm laser but did not seem to be present in measurements involving the shorter wavelength solid state lasers. The detector was then profiled by taking power measurements at equally spaced 0.0625 cm increments creating a 1.2 cm$^2$ grid, shown in Figure \ref{profile}.  The photodetector has a circular effective measuring area with a radius of 0.5 cm.  It is clear from these data that the 543 and 632 nm lasers record large fluctuations at the edges of the detector, with the maximum located at the top edge, while the solid state lasers remain consistent across the detector surface.  This suggests that there may be some internal reflection inside the photodetector that becomes pronounced above a certain wavelength between 440 and 543 nm.  The actual detection film is located 1.5 cm inside the detector casing.  This design creates a ``lip'' that might catch light reflected off the film and direct it back into the detector. To investigate the possibility of internal reflection, two sets of measurements were taken using the 543 nm laser, one at the center of the detector and one at the upper edge, with a varying incident angle, shown in Figure \ref{angle}.  At the center of the detector the recorded power is lower and stable, $\approx$418 $\mu$W.  On the upper edge readings are much larger, even at normal incidence, $\approx$1475 $\mu$W, and rise linearly as the angle of incidence is increased.  At $\approx$ 2.25$^{\circ}$ the power readings level off giving a reading of $\approx$1775 $\mu$W corresponding to the angle needed for maximum internal reflection.  In order to take accurate measurements at the longer wavelengths it is important that the incoming light be normally incident on the center of the detector.  This can be difficult when passing the beam through the acrylic samples because small deviations from normal incidence at the acrylic/air interface can cause the light to enter the detector at undesirable locations and angles.  To reduce systematic error a convex lens was installed before the photodetector.  This effectively restricts the size of the beam, directs it to a single reproducible point, and aids in the aiming process which reduces the systematic error by restricting the area where the beam interacts with the detector.  

\subsection{Method}
In order to calculate the transmittance and attenuation length of light through acrylic this experiment compared the power of incoming monochromatic light, supplied by a laser, in air to that of monochromatic light propagating through an intermediate acrylic medium.  To begin testing, each acrylic rod was placed on a set of mounted rollers that keep the acrylic in a stable position while allowing it to rotate about its axis of symmetry. This was only done with cylindrical acrylic samples.  For the rectangular acrylic samples blocks were used to achieve the desired height but the sample was unable to rotate.  The two machined surfaces were then polished a final time using a three step acrylic polish developed by NOVUS \cite{novus}.  The first laser was then mounted, leveled, and aligned with the power meter in such a way that the beam passed through the entire acrylic sample, with normal incidence on the two polished surfaces, and no internal reflection.  The beam was incident on the acrylic at a location halfway between the center and the outer edge of the surface so that successive rotations of the samples by 90$^\circ$ provided four independent (unique) transmission measurements.  Since the rectangular acrylic samples could not rotate, unique sections were marked out and the entire sample was moved so that the laser passed though them.  All of these components were then covered by a black optical cloth to block stray light.  In general, a dim room light was left on to ease the testing process.  It was found that measurements done with the room lights on but the black cloth present fluctuated by $<1\%$ compared to those done in complete darkness.  A measurement of the power was then taken without the presence of acrylic by sliding the entire roller mount from the beam path.  Moving the acrylic back into the beam path another measurement of the power was taken and the acrylic sample was rotated 90$^\circ$.  This process was then repeated without acrylic and for four unique paths through the sample.  Upon the completion of subsequent runs, a laser of a different wavelength was mounted, aligned, and the process was repeated.

\section{Calculating Transmittance Through Acrylic}
As an intermediate step to calculating the attenuation length, we calculated the transmittance of light through each acrylic sample.  The simplest definition of transmittance, $T$, is the ratio of the intensities of incoming light in air $I_{air}$ to that of light passing through a medium $I_{acrylic}$:

\begin{equation}
T= \frac{I_{acrylic}}{I_{air}} \label{trans_simple}
\end{equation} 

$I_{acrylic}$ can be defined in terms of $I_{air}$, the coefficient of attenuation $\alpha$, and the distance it travels through the medium $d$:

\begin{equation}
I_{acrylic}=I_{air} e^{-\alpha d} \label{inten}
\end{equation}

and substituting Equation \ref{trans_simple} into \ref{inten}

\begin{equation}
T=e^{-\alpha d} \label{trans_exp}
\end{equation} 

\subsection{Reflection at Acrylic Faces}

A fraction of the incoming light is expected to be lost, due to reflection.  This reflection is dependent on the polarization of the incoming light.  For light with an electric field perpendicular to the plane of incidence (s polarization) the reflection coefficient $R_s$ is given by:

\begin{equation}
R_s=\left(\frac{n_1\cos{\theta}-n_2\sqrt{1-(\frac{n_1}{n_2}\sin{\theta})^2}}{n_1\cos{\theta}+n_2\sqrt{1-(\frac{n_1}{n_2}\sin{\theta})^2}}\right)^2
\end{equation}

\noindent where $n_1$ is the index of refraction of air ($n_1=1$), $n_2$ is the index of refraction of acrylic, and $\theta$ is the angle of incidence.  For light with an electric field parallel to the plane of incidence (p polarization) the reflection coefficient $R_p$ is given by:

\begin{equation}
R_p=\left(\frac{n_1\sqrt{1-(\frac{n_1}{n_2}\sin{\theta})^2}-n_2\cos{\theta}}{n_1\sqrt{1-(\frac{n_1}{n_2}\sin{\theta})^2}+n_2\cos{\theta}}\right)^2
\end{equation}

The lasers used in this experiment all emitted unpolarized light which is by definition equal parts p and s polarized light

\begin{equation}
R=\frac{R_s+R_P}{2}
\end{equation}

or

\begin{equation}
R=\frac{1}{2}\left(\frac{n_1\cos{\theta}-n_2\sqrt{1-(\frac{n_1}{n_2}\sin{\theta})^2}}{n_1\cos{\theta}+n_2\sqrt{1-(\frac{n_1}{n_2}\sin{\theta})^2}}\right)^2+\frac{1}{2}\left(\frac{n_1\sqrt{1-(\frac{n_1}{n_2}\sin{\theta})^2}-n_2\cos{\theta}}{n_1\sqrt{1-(\frac{n_1}{n_2}\sin{\theta})^2}+n_2\cos{\theta}}\right)^2
\end{equation}

Using the index of refraction value of acrylic ($n_2=1.5$) we calculate the loss of light due to reflection to be $\approx4\%$ using $\theta=0$.  This agrees well with the measurements of the reflected light at different wavelengths while trying to minimize the angle of incidence.  Light is reflected at both acrylic surfaces giving a total loss of light due to reflection of $\approx8\%$.  Multiple reflections were calculated to be a negligible effect.  Since the transmittance is related to the reflection by

\begin{equation}
T=1-R
\end{equation}

the maximum amount of light transmitted through two surfaces is $T=92\%$.  We can then modify Equation \ref{trans_simple} to get a final equation for calculating the transmittance

\begin{equation}
T=\frac{I_{acrylic}}{(0.92)I_{air}}
\end{equation}

\subsection{Transmittance with Respect to Position}

One variable that affects transmittance measurements are surface effects or inconsistencies in the manufacturing of the acrylic samples.  To examine the transmittance dependence on orientation the experiment was designed so that each acrylic sample could be measured at different positions.  Once a set of four positional measurements were completed and the expected value of the transmittance calculated for each position, the difference between the position with the maximum value and the position with the minimum value was calculated and are presented in Table \ref{diff}.  The table shows that the largest difference in transmittance is $6.2\pm2.4\%$, but in general differences are much lower with several measuring at $<1\%$.  The largest variance in individual measurements is found in the higher wavelengths but greater differences as a function of orientation appear at lower wavelengths.    

\begin{table} \centering
\begin{tabular}{|c|c|c|c|c|c|} \hline\hline
{\em Sample} & \multicolumn{5}{c|}{\em Wavelength (nm)} \\ \hline
 & 375 &405 &440 &543 &632 \\ \hline
 & \multicolumn{5}{c|}{\em Difference (\%)} \\ \hline 
	Tecacryl \#1 & $0\pm0.1$ & $0.4\pm0.5$ & $0.7\pm2.0$ & $0.7\pm1.8$ & $2.8\pm2.1$ \\
	Tecacryl \#2 & $0\pm0.1$ & $0.7\pm0.1$ & $4.3\pm0.8$ & $2.3\pm1.2$ & $1.2\pm2.4$ \\
	Tecacryl \#3 & $1.5\pm0.8$ & $1.3\pm0.4$ & $1.0\pm1.0$ & $3.0\pm1.1$ & $1.3\pm2.5$ \\
	Tecacryl \#4 & $1.8\pm1.1$ & $1.2\pm1.8$ & $1.1\pm1.2$ & $2.0\pm1.7$ & $0.8\pm3.8$ \\
	RPT \#1 & $3.0\pm0.4$ & $2.8\pm0.3$ & $0.8\pm1.0$ & $1.9\pm1.7$ & $0.5\pm3.4$ \\
	RPT \#2 & $3.8\pm0.1$ & $3.2\pm0.9$ & $2.2\pm1.1$ & $1.4\pm1.6$ & $1.1\pm2.7$ \\
	RPT \#3 & $2.9\pm0.1$ & $4.1\pm0.3$ & $2.4\pm0.8$ & $1.5\pm0.6$ & $1.1\pm2.7$ \\
	RPT \#4 & $0\pm0.1$ & $0.2\pm0.1$ & $2.1\pm0.8$ & $1.6\pm1.8$ & $0.5\pm2.2$ \\
	Spartech \#1 & $1.5\pm0.9$ & $4.2\pm1.8$ & $3.9\pm1.1$ & $1.1\pm1.1$ & $1.8\pm1.8$ \\
	Spartech \#2 & $0.9\pm0.9$ & $1.1\pm2.5$ & $1.0\pm1.4$ & $0.9\pm1.4$ & $0.9\pm3.5$ \\
  Spartech \#3 (short side) & $1.0\pm0.6$ & $1.1\pm1.1$ & $1.0\pm1.8$ & $1.9\pm1.1$ & $0.3\pm2.3$ \\
  Spartech \#3 (long side) & $0\pm0.1$ & $0.5\pm0.3$ & $0.6\pm0.7$ & $1.4\pm1.4$ & $1.3\pm2.7$ \\
	Spartech \#4 & $0\pm0.1$ & $0.2\pm0.1$ & $6.2\pm2.4$ & $1.0\pm1.2$ & $1.3\pm4.1$ \\ 
	McMaster (Spartech) \#1 & $0.5\pm2.3$ & $1.4\pm0.8$ & $0.4\pm0.9$ & $0.1\pm0.9$ & $0.4\pm2.9$ \\
	McMaster (US Cast) \#2 & $0.1\pm0.3$ & $0.7\pm0.4$ & $0.1\pm2.0$ & $0.2\pm0.9$ & $0.4\pm1.6$ \\
\hline\hline
\end{tabular}
\caption{The difference between the minimum and maximum transmittance values reordered for the set of four unique positions of the acrylic sample.  All values are measured as \% of light transmitted.} 
\label{diff}
\end{table} 

\subsection{Transmittance Independent of Position}
\begin{figure}\centering
\includegraphics[width=0.49\linewidth]{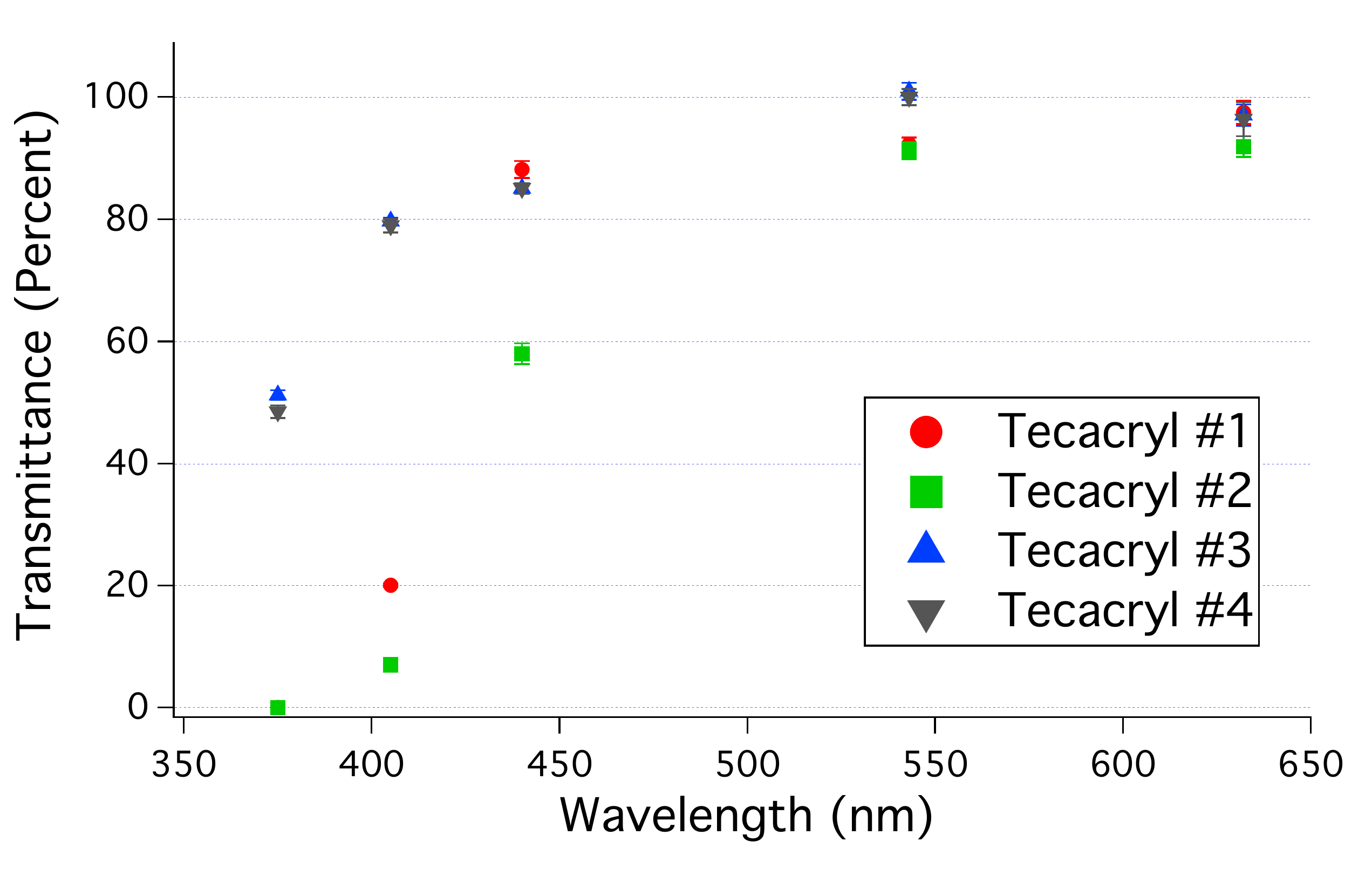}
\includegraphics[width=0.49\linewidth]{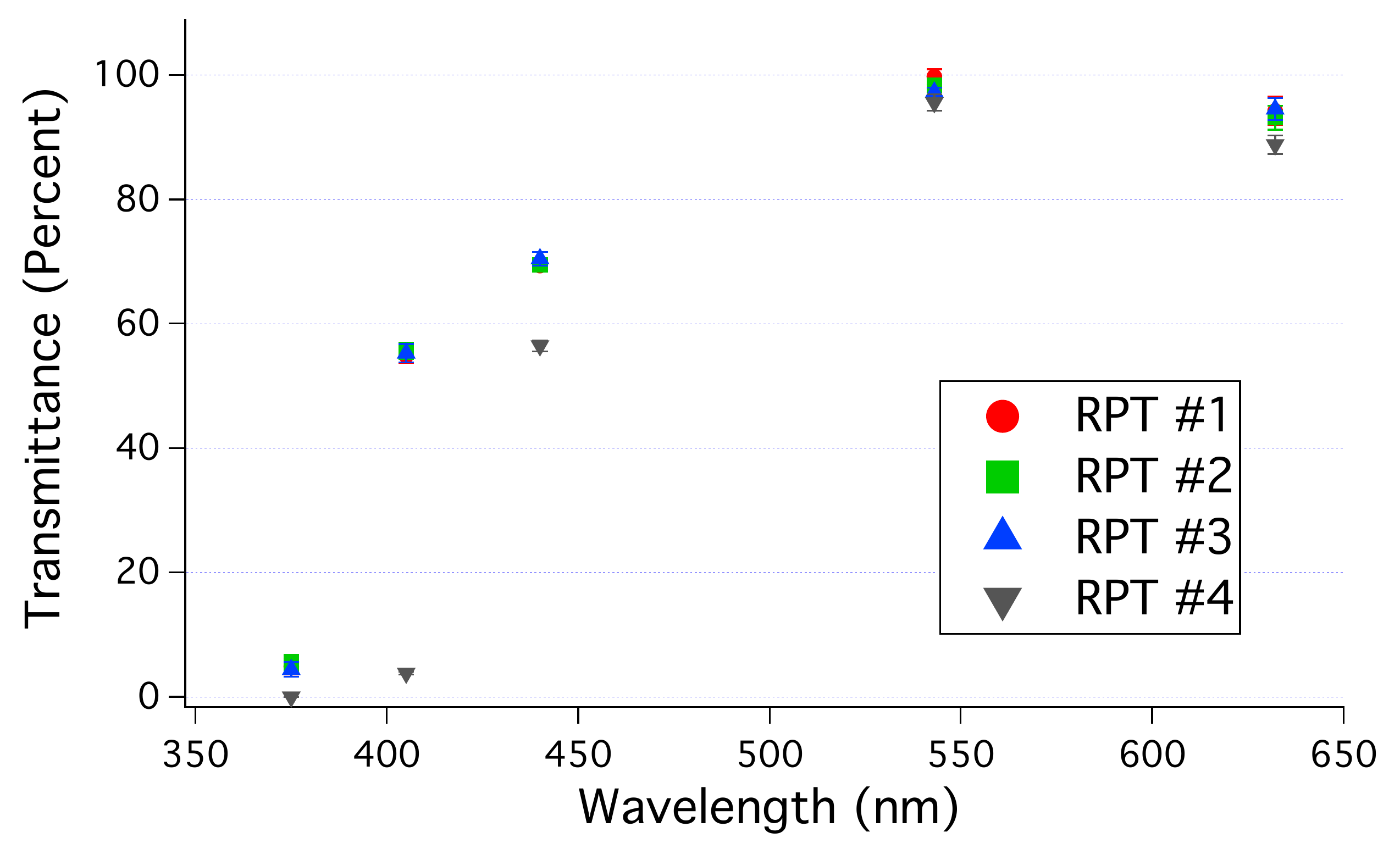}
\includegraphics[width=0.49\linewidth]{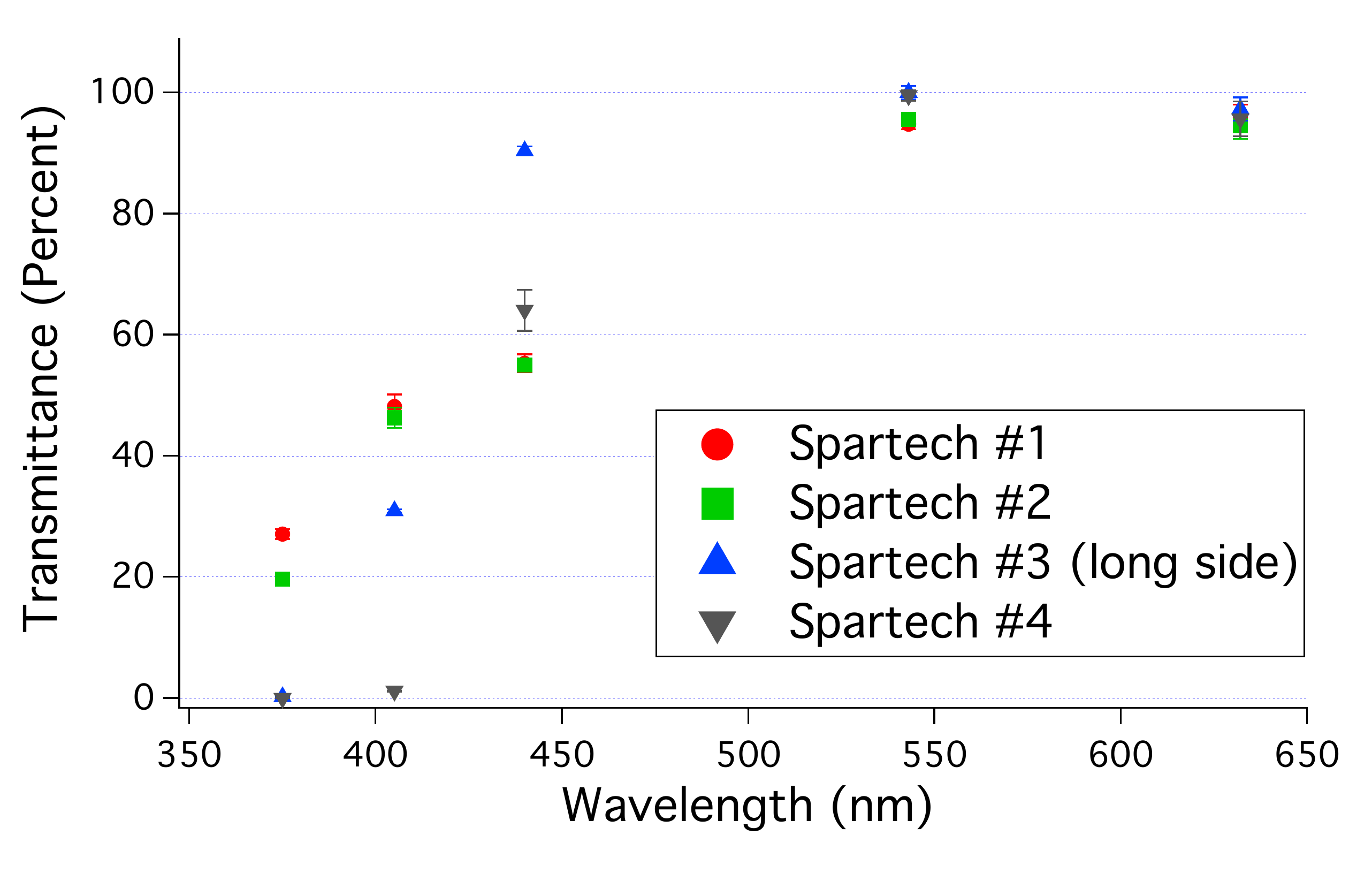}
\includegraphics[width=0.49\linewidth]{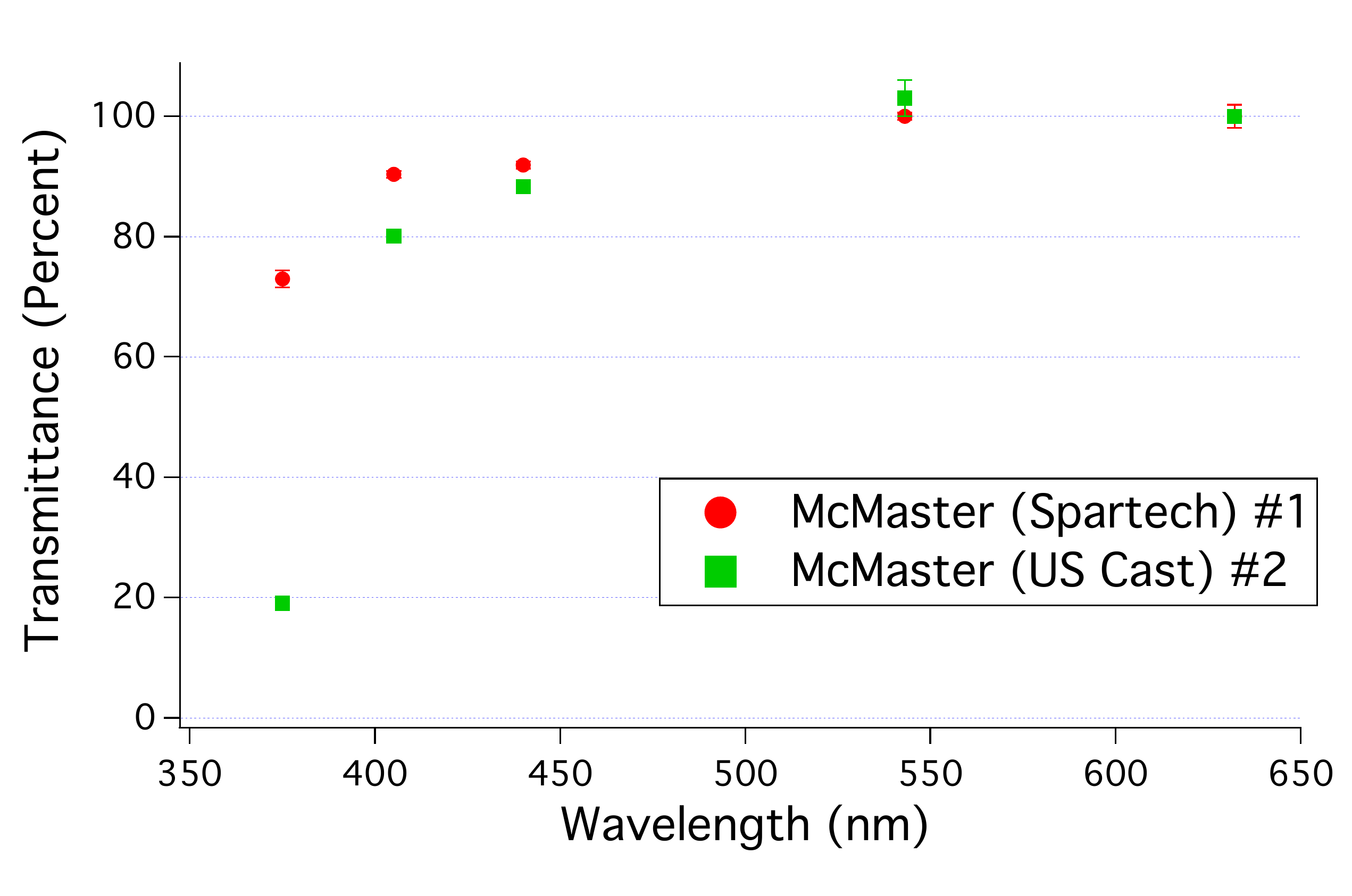}
\caption{Plots of transmittance as a function of wavelength for acrylic samples.  Plotted points represent the mean value for the entire set of measurements over all orientations.}
\label{transplot}
\end{figure}

Given the low dependence on orientation of the acrylic samples used it is reasonable to calculate the transmittance as a set of all positions.  Here differences based on the position of the acrylic sample are included in the overall variance of the transmittance.  The result for each sample can be found in Table \ref{comb}.  It is important to remember that the transmittance is only useful to observe the general dependence on wavelength of an individual piece of acrylic or to compare samples with equal lengths.  Plots comparing the transmittance of the acrylic samples from each manufacturer are shown in Figure \ref{transplot}.  From the table and accompanying plots it is clear that all the acrylic samples have the same general trend.  At larger wavelengths (543 and 632 nm) there is almost total  transmittance even in relatively long samples such as Spartech $\#4$.  The transmittance then starts to diminish somewhere between 543 and 440 nm and rapidly decreases as we approach 375 nm.  As expected, response at lower wavelengths is a function of the sample type (UVA vs. UVT).  The uncertainty on the transmittance includes the variation between measurements at different positions, the uncertainty in the power meter reading, and the stability of the lasers during the measurement period.

\begin{table} \centering
\begin{tabular}{|c|c|c|c|c|c|} \hline\hline
{\em Sample} & \multicolumn{5}{c|}{\em Wavelength (nm)} \\ \hline
 & 375 &405 &440 &543 &632 \\ \hline
 & \multicolumn{5}{c|}{\em Transmittance (\%)} \\ \hline
	Tecacryl \#1 & $0.1\pm0.3$ & $20.1\pm0.3$ & $88.2\pm1.4$ & $92.3\pm1.1$ & $97.5\pm1.9$  \\
	Tecacryl \#2 & $0.1\pm0.0$ & $7.1\pm0.3$ & $58.0\pm1.7$ & $91.3\pm1.4$ & $91.9\pm1.7$ \\
	Tecacryl \#3 & $51.2\pm0.8$ & $79.7\pm0.6$ & $85.1\pm0.9$ & $100.0\pm1.4$ & $97.1\pm1.8$ \\
	Tecacryl \#4 & $48.5\pm1.0$ & $79.0\pm1.1$ & $82.9\pm1.0$ & $100.0\pm1.3$ & $96.4\pm2.8$ \\
	RPT \#1 & $4.9\pm1.3$ & $54.9\pm1.1$ & $69.4\pm0.7$ & $99.8\pm1.2$ & $94.3\pm2.3$ \\
	RPT \#2 & $5.4\pm1.4$ & $55.7\pm1.3$ & $69.5\pm1.1$ & $98.4\pm1.1$ & $93.2\pm1.9$ \\
	RPT \#3 & $4.4\pm1.2$ & $55.3\pm1.5$ & $70.5\pm1.1$ & $97.3\pm0.7$ & $94.6\pm1.8$ \\
	RPT \#4 & $0.0\pm0.0$ & $3.8\pm0.2$ & $56.5\pm0.9$ & $95.6\pm1.3$ & $88.8\pm1.5$ \\
	Spartech \#1 & $27.1\pm0.8$ & $48.2\pm2.0$ & $55.3\pm1.5$ & $94.8\pm0.8$ & $95.9\pm2.1$ \\
	Spartech \#2 & $19.7\pm0.6$ & $46.3\pm1.7$ & $55.0\pm0.9$ & $95.6\pm0.9$ & $94.6\pm2.2$ \\
	Spartech \#3 (short side) & $22.5\pm0.6$ & $64.8\pm0.8$ & $95.9\pm1.5$ & $100.0\pm1.0$ & $99.6\pm1.5$ \\
	Spartech \#3 (long side) & $0.2\pm0.0$ & $30.9\pm0.3$ & $90.4\pm0.7$ & $100.0\pm1.1$ & $97.3\pm1.9$ \\
	Spartech \#4 & $0.0\pm0.0$ & $1.2\pm0.1$ & $64.1\pm3.4$ & $99.5\pm0.9$ & $95.7\pm2.9$ \\
	McMaster (Spartech) \#1 & $73.0\pm1.4$ & $90.4\pm0.6$ & $91.9\pm0.6$ & $100.0\pm0.6$ & $100.0\pm1.9$ \\
	McMaster (US Cast) \#2 & $19.1\pm0.2$ & $80.1\pm0.3 $ & $88.3\pm1.1$ & $100.0\pm0.6$ & $100.0\pm1.0$ \\
\hline\hline
\end{tabular}
\caption{Transmittance values for all acrylic samples combining the data from all positions.} 
\label{comb}
\end{table} 

\section{Calculating the Attenuation Length}
Since the transmittance is dependent on the length of the acrylic sample it is not useful to compare across all acrylic samples.  To remove the dependence on sample length it is necessary to calculate the attenuation length.  Attenuation length is defined as the length of material needed to reduce the incoming beam Intensity to $\frac{1}{e}$ or to have about 63\% of the incoming particles stopped.  Recall from Equation \ref{trans_exp}:

\begin{equation}
T(d)=e^{-\alpha d} \label{atteq}
\end{equation}
 	
\noindent where $\alpha$ is the attenuation coefficient

\begin{equation}
\alpha =\frac{1}{\lambda} \label{lambda}
\end{equation}

\noindent and $\lambda$ is the attenuation length.  By substituting Equation \ref{lambda} into Equation \ref{atteq} and solving for the attenuation length we obtain:

\begin{equation}
\lambda =\frac{d}{-\log{[T(d)]}}
\end{equation}

\begin{figure}\centering
\includegraphics[width=.49\linewidth]{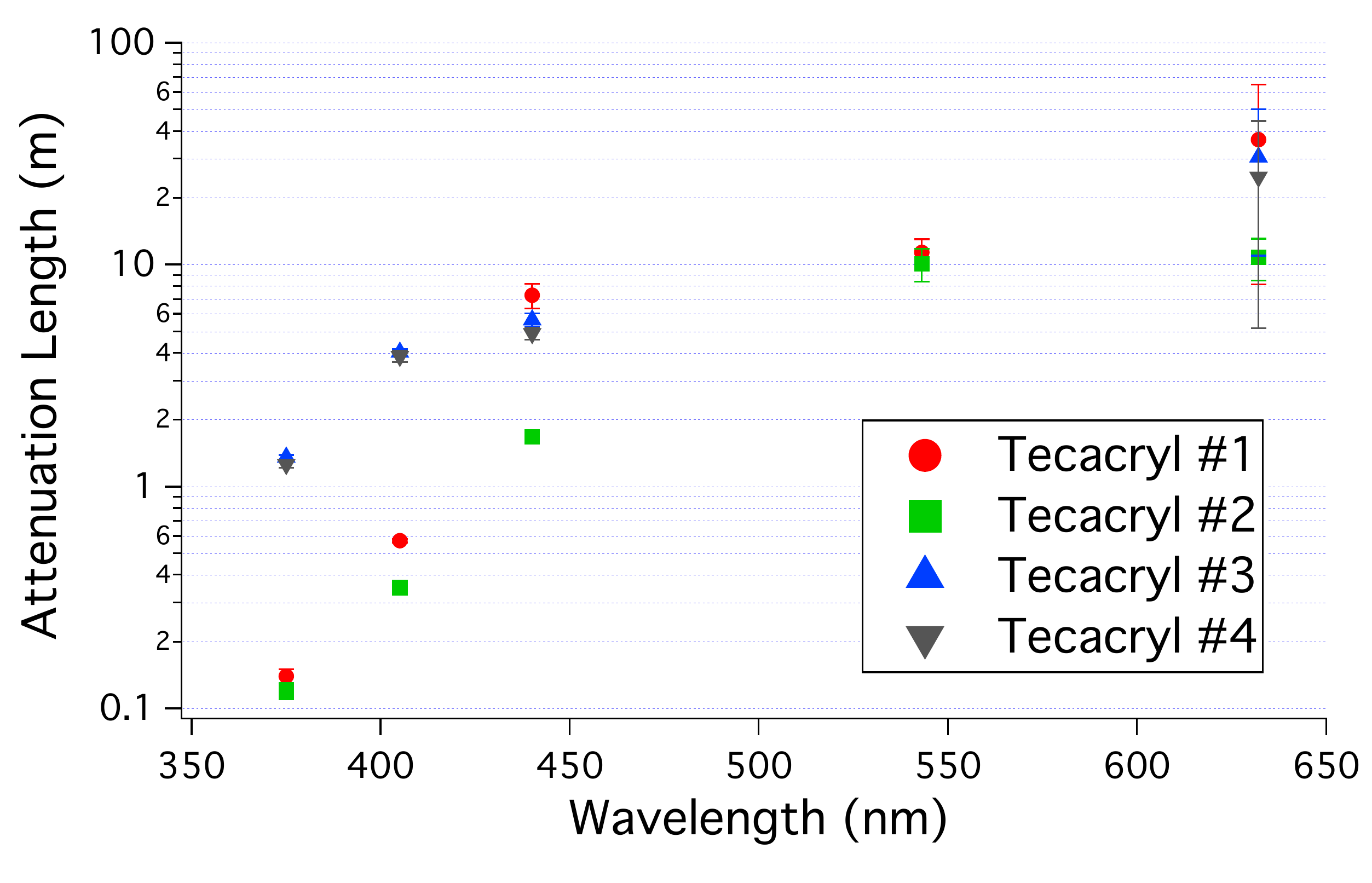}
\includegraphics[width=.49\linewidth]{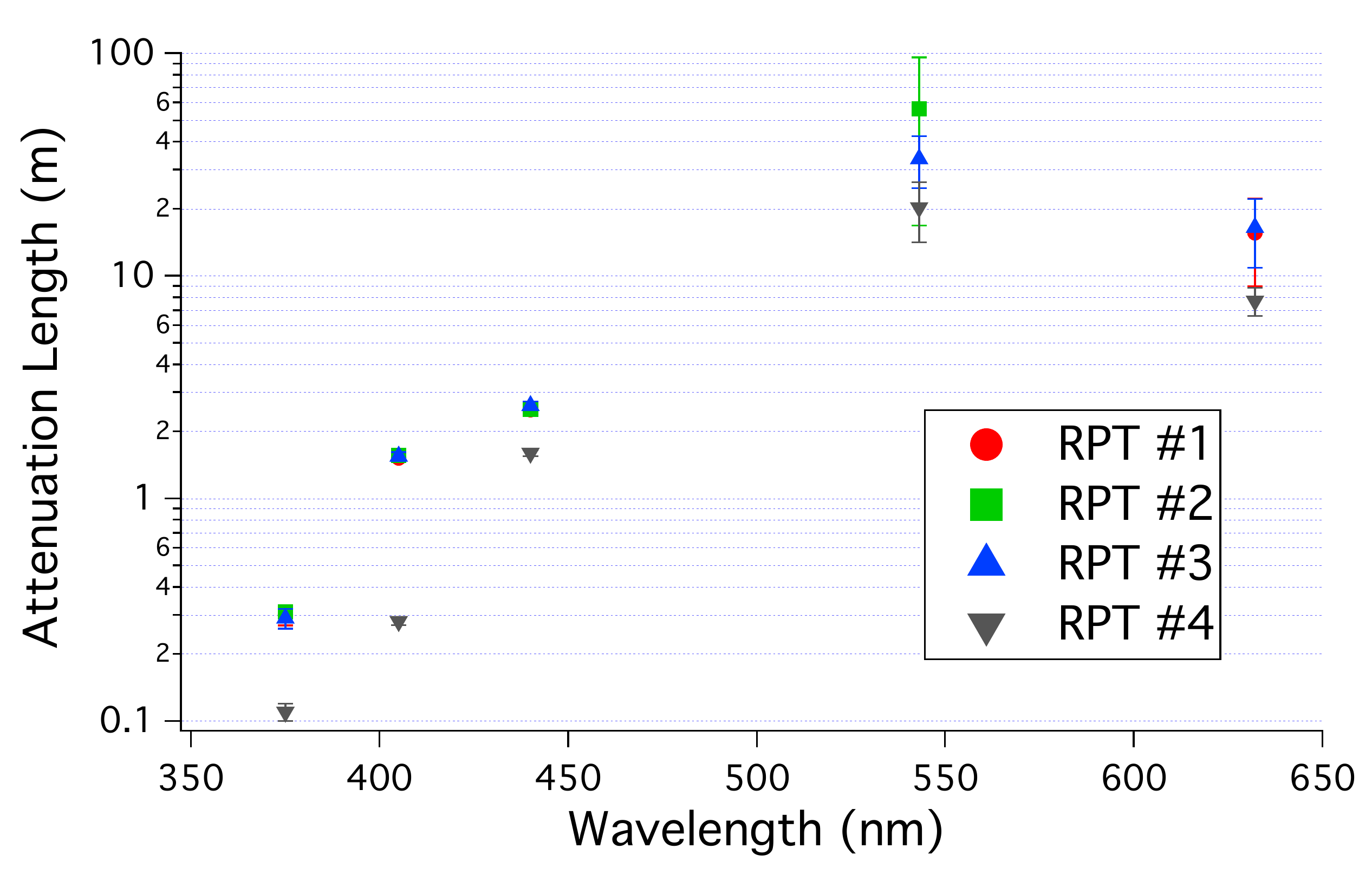}
\includegraphics[width=.49\linewidth]{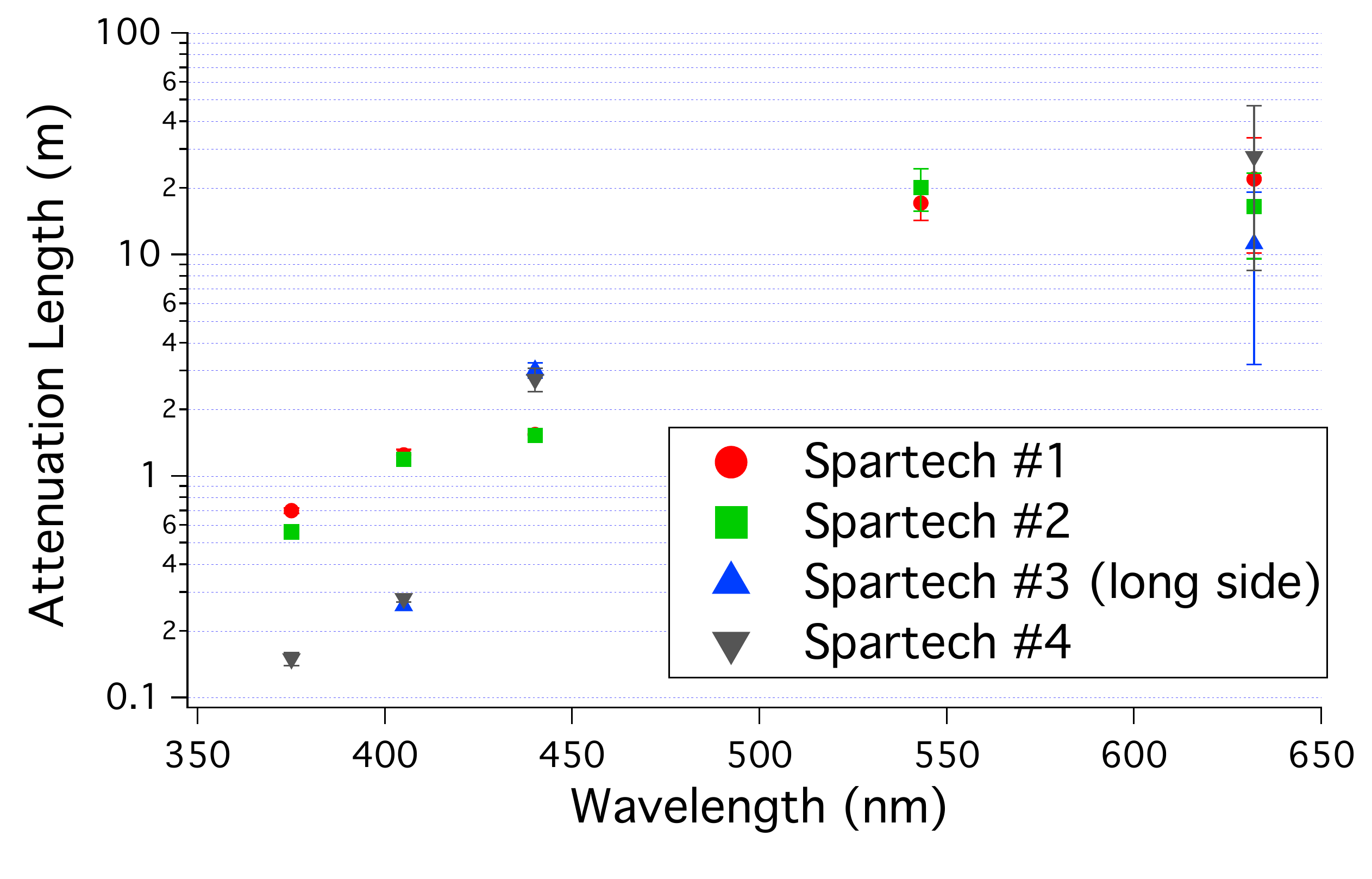}
\includegraphics[width=.49\linewidth]{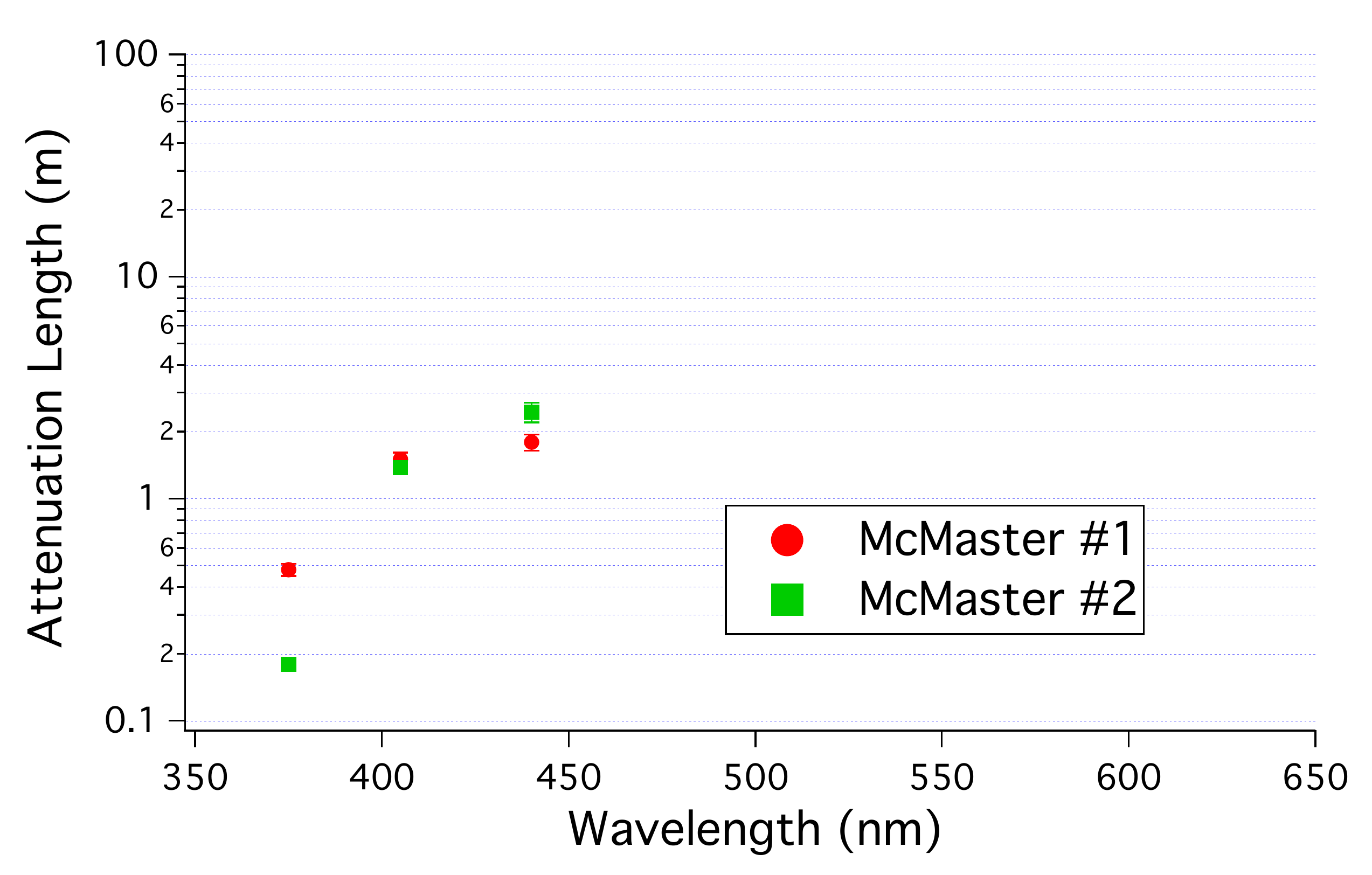}
\caption{Plots of the attenuation length for all acrylic samples grouped by manufacturer.  Values for wavelengths where the attenuation length approached $\infty$ are not shown.}
\label{attplot}
\end{figure}

The attenuation length is independent of the distance which makes it possible to compare all the acrylic samples regardless of their dimensions.  Our method using large samples of acrylic has improved the error of measuring large attenuation lengths that can be quite large when measuring small samples in spectrophotometers.  However, even our method results in large errors when the optical attenuations lengths exceed several meters as is the case for many of the measurements at the highest wavelengths (543 and 632 nm) and at 440 nm for shorter acrylic samples.  Acrylic samples can now have transmittance of near 100\% at 543 and 632 nm even for samples as large as 121.92 cm.  Therefore for light guides with lengths that are significantly shorter, the attenuation length can be assumed to be very large and the only light lost must be due to reflection at the surfaces.  Results for the attenuation can be found in Table \ref{atttable} and plots at selected wavelengths can be found in Figure \ref{attplot}.      

\begin{table} \centering
\begin{tabular}{|c|c|c|c|c|c|} \hline\hline
{\em Sample} & \multicolumn{5}{c|}{\em Wavelength (nm)} \\ \hline
 & 375 &405 &440 &543 &632 \\ \hline
 & \multicolumn{5}{c|}{\em Attenuation Length (m)} \\ \hline 
	Tecacryl \#1 & $0.14\pm0.01$ & $0.57\pm0.01$ & $7.30\pm0.93$ & $11.4\pm1.7$ & $36.7\pm28.5$ \\
	Tecacryl \#2 & $0.12\pm0.00$ & $0.35\pm0.01$ & $1.68\pm0.09$ & $10.1\pm1.7$ & $10.8\pm2.3$ \\
	Tecacryl \#3 & $1.36\pm0.03$ & $4.03\pm0.14$ & $5.65\pm0.38$ & -- & $30.6\pm19.6$ \\
	Tecacryl \#4 & $1.26\pm0.04$ & $3.87\pm0.22$ & $4.89\pm0.30$ & -- & $24.9\pm19.7$ \\
	RPT \#1 & $0.30\pm0.03$ & $1.52\pm0.05$ & $2.50\pm0.07$ & $492\pm3270$ & $15.6\pm6.61$ \\
	RPT \#2 & $0.31\pm0.02$ & $1.56\pm0.06$ & $2.51\pm0.11$ & $56.2\pm39.4$ & $2.9\pm3.7$ \\
	RPT \#3 & $0.29\pm.03$ & $1.55\pm0.07$ & $2.61\pm0.11$ & $33.6\pm8.80$ & $16.5\pm5.6$ \\
	RPT \#4 & $0.11\pm0.00$ & $0.28\pm0.01$ & $1.60\pm0.05$ & $20.3\pm6.1$ & $7.7\pm1.1$ \\
	Spartech \#1 & $0.70\pm0.02$ & $1.25\pm0.07$ & $1.54\pm0.07$ & $17.1\pm2.8$ & $22.0\pm11.8$ \\
	Spartech \#2 & $0.56\pm0.01$ & $1.19\pm0.06$ & $1.53\pm0.04$ & $20.1\pm4.4$ & $16.5\pm6.9$ \\
  Spartech \#3 (short side) & $0.07\pm0.00$ & $0.23\pm0.01$ & $2.41\pm0.92$ & -- & $24.9\pm95.0$ \\
  Spartech \#3 (long side) & $0.05\pm0.00$ & $0.26\pm0.00$ & $3.01\pm0.23$ & -- & $11.2\pm8.0$ \\
	Spartech \#4 & $0.15\pm0.00$ & $0.28\pm0.01$ & $2.74\pm0.33$ & $230\pm379$ & $27.8\pm19.3$ \\ 
	McMaster (Spartech \#1 & $0.48\pm0.03$ & $1.51\pm0.11$ & $1.80\pm0.15$ & -- & -- \\
	McMaster (US Cast) \#2 & $0.18\pm0.00$ & $1.38\pm0.03$ & $2.46\pm0.25$ & -- & -- \\
\hline\hline
\end{tabular}
\caption{Table of attenuation lengths for all acrylic samples.  Entries marked as `--' have a calculated transmittance of 100\%.  The function used for calculating the attenuation length approaches $\infty$ as transmittance approaches 100$\%$.  Since the attenuation length is independent of the samples dimension this table can be used to compare samples of acrylic with different lengths.} 
\label{atttable}
\end{table} 

The results show great variability in optical attenuation between manufacturers but reasonable repeatability by each manufacturer.  For overall optical attenuation for UVT acrylic, the Tecacryl from Plastifab proved to be superior for the samples with the proper additive mixture (samples \#3 and \#4).

\section{Conclusion}
Acrylic can be implemented in many detectors as an inexpensive light guide.  It is essential though that the acrylic used allows for the maximum amount of light to be transmitted.  The amount of transmitted light is dependent on the wavelength of the light and the distance that light must travel through the acrylic.  Using the attenuation length one can compare samples from various acrylic manufacturers.

For the MiniCLEAN experiment, the acrylic must have good transmittance at wavelengths above 420 nm to be well matched to the re-emmission spectrum of the wavelength shifter.  It also must be radioactively clean.  Based on these requirements and the ability to work with the manufacturer on cleanliness during the processing, the MiniCLEAN detector utilizes UVA acrylic sheets cast by Spartech (Spartech \#3 and \#4).  This acrylic was chosen since the final thickness of acrylic used in MiniCLEAN was only 10 cm and the Spartech acrylic was available in cell cast sheets of that thickness making it more economical than single cylindrical castings.  Spartech was also willing to follow rigorous quality assurance procedures (such as verification of clean molds and limiting dust exposure) and utilize monomer immediately after its production to reduce contamination by radon gas naturally present in the air.  However, other experiments with different requirements and cleanliness concerns will find several excellent acrylic products available on the market today. 
 
\acknowledgments
We gratefully acknowledge the University of New Mexico Physics and Astronomy machine shop for their assistance in machining and polishing the acrylic samples and constructing the roller apparatus.  This work was funded by the Los Alamos National Laboratory's Laboratory Directed Research and Development program and the Department of Energy Office of Science High Energy Physics program.

\end{document}